\documentclass[10pt,pra,aps,twocolumn,superscriptaddress,floatfix,notitlepage,nofootinbib]{revtex4-2}
\usepackage[latin1]{inputenc}
\usepackage{hyperref}
\hypersetup{
    colorlinks=true,
    linkcolor=blue,
    filecolor=magenta,      
    urlcolor=cyan,
    pdftitle={Detecting Acoherence in Radiation Fields},
    pdfpagemode=FullScreen,
    }
\usepackage{mathrsfs,dsfont}
\usepackage{amsmath}
\usepackage{amsfonts,natbib}
\usepackage{amssymb}
\usepackage{bm}
\usepackage[pdftex]{graphicx}

\usepackage{array}

\usepackage{textpos}
\usepackage{bbold}
\usepackage{adjustbox}

\usepackage{amsmath}
\usepackage{graphicx}
\begin{document}
\title{Testing the coherent-state description of radiation fields}
\author{Sreenath K. Manikandan}
\email{sreenath.k.manikandan@su.se}
\affiliation{Nordita, Stockholm University and KTH Royal Institute of Technology, Hannes Alfv\'{e}ns v\"{a}g 12, SE-106 91 Stockholm, Sweden}
\author{Frank Wilczek}
\affiliation{Center for Theoretical Physics, Massachusetts Institute of Technology, Cambridge, Massachusetts 02139, USA}
\affiliation{T. D. Lee Institute, Shanghai 201210, China}
\affiliation{Wilczek Quantum Center, Department of Physics and Astronomy, Shanghai Jiao Tong University, Shanghai 200240, China }
\affiliation{Department of Physics and Origins Project, Arizona State University, Tempe, Arizona 25287, USA}
\affiliation{Department of Physics, Stockholm University, AlbaNova University Center, 106 91 Stockholm, Sweden}
\affiliation{Nordita, Stockholm University and KTH Royal Institute of Technology, Hannes Alfv\'{e}ns v\"{a}g 12, SE-106 91 Stockholm, Sweden}
\date{\today}
 \begin{abstract}
  We propose simple quantitative criteria, based on counting statistics in resonant harmonic detectors, that probe the quantum mechanical character of radiation fields.  They provide, in particular, practical means to test the null hypothesis that a given field is ``maximally classical'', i.e., accurately described by a coherent state.  We suggest circumstances in which that hypothesis plausibly fails, notably including gravitational radiation involving non-linear or stochastic sourcing.
\end{abstract}

\begin{textblock*}{5cm}(12cm,-15cm)
\fbox{\footnotesize MIT-CTP/5485}
\end{textblock*}

\maketitle

\section{Introduction}
Assuming the validity of quantum theory, in principle all radiation fields should be described as quantum-mechanical states (or density matrices).  But alternative descriptions can often be adequate and, when they are, easier to use.  In particular, a description based on classical physics can capture the properties of a class of states known as coherent states, that provide an adequate approximation in many situations of interest.   Coherent states, in the relevant sense, are eigenstates of the field amplitude operator, $a$ . They can be shown to arise when the radiation field is generated by linear coupling to a classical source.  Coherent states have zero quantum mechanical variance (noise) for the field amplitude operators,
    \begin{equation}
        \langle \alpha|(\Delta a)^2|\alpha\rangle = \langle \alpha|(\Delta a^{\dagger})^2|\alpha\rangle = 0.  
    \end{equation} 
      They are widely used in quantum optics, and they are implicitly assumed whenever radiation is treated as a classical field 
        (formally via
    the optical equivalence theorem~\cite{Sudharshan,Mandel_Wolf_1995,Glauber}).  

Deviation of radiation fields from coherent states demonstrate the inadequacy of a classical description, and can reveal important information about the radiation field and its sources.  It is therefore desirable to identify simple experimental tests that discriminate the quantum state of the field from a coherent state.  That is our goal here.

Since the word ``coherence'' is often with different meanings and different antonyms -- {\it e.g.}, broadly, in the statistical characterization of disorder within classical optical fields (opposite: partial coherence) or quantum density matrices (opposite: partial decoherence) -- here we will mean by an {\it acoherent\/} state a state that is not a coherent state in the precise sense mentioned above, and by {\it acoherence\/} deviation from the generic behavior of such coherent states.

\section{Measurement Model}
We consider two resonantly coupled quantum harmonic oscillators. The interaction Hamiltonian between the two oscillators is assumed to be
\begin{equation}\label{interaction_Hamiltonian}
    V_I(t) = \hbar\sqrt{\gamma_0} [d(t)a^\dagger+d^{\dagger}(t)a].
\end{equation} 
With   the $a$ mode   representing a mode of the radiation field and the   $d$ mode   the detector, this set-up provides a useful model for practical detection schemes.  $\sqrt{\gamma_0}$ is the effective coupling constant between the radiation field and the detector.  It arises in the rotating wave, quasi-monochromatic    approximation for the mode of the field (assumed, for simplicity, unique) at the detector's resonant frequency.  For a near resonant continuous radiation field incident on the detector, after making the rotating wave approximation and incorporating the density of modes,
$\gamma_0$ can also be identified as the spontaneous emission rate of the detector~\cite{jordan_anatomy_2016,lewalle_measuring_2020,gross_qubit_2018} as shown in Appendix.~\ref{graviton}. Similar considerations also arise for two-state detectors, i.e. qubits ~\cite{gross_qubit_2018}.  

We want to evaluate the time evolution operator for an interval of observation, in the form
\begin{equation}
    U_I = e^{-i\sqrt{\gamma_0}\int_t^{t+\Delta t} [d(t')a^\dagger+d^{\dagger}(t')a]dt'}.
\end{equation}
where $[d(t'),d^{\dagger}(t'')]=\delta (t'-t'')$.  
We have absorbed any residual time-dependence to the detector mode $d(t)$ in Eq.~\eqref{interaction_Hamiltonian}, and the delta function above implies that the detector modes at different times are independent. Going forward,  
it is convenient to introduce operators $b=\frac{1}{\sqrt{\Delta t}}\int_{t}^{t+\Delta t} d(t')dt'$ satisfying $[b,b^{\dagger}]=1$.
Using these, we can write the evolution operator as
\begin{equation}
    U_I = e^{-i\sqrt{\gamma_0}\int_t^{t+\Delta t} [d(t')a^\dagger+d^{\dagger}(t')a]dt'} =e^{-i\sqrt{\gamma_0\Delta t }(ba^\dagger+b^{\dagger}a)}.  
\end{equation}
Involving the effective Hamiltonian  
$H_{I} \Delta t/\hbar \equiv \sqrt{\gamma_0 \Delta t}(a^{\dagger} b+b^{\dagger}a )\equiv \kappa (a^{\dagger} b+b^{\dagger}a )$,
we can evaluate the probabilities $P_n$ that the detector, initialized in its ground state, is excited to the $n^{\rm th}$ level. To order $\kappa^6$, in field state $|s \rangle$, we find:
\begin{eqnarray}\label{po}
    &P_0& ~=~ 1 - \kappa^2 \langle s|a^\dagger a|s\rangle + \kappa^4 \langle s|\frac{1}{6} a^{\dagger 2} a^2 + \frac{1}{3} (a^\dagger a)^2 | s\rangle\nonumber  \\
   ~&-&~ \kappa^6\langle s|  \frac{17}{360} a^\dagger a a^{\dagger 2} a^2  +  \frac{17}{360} a^{\dagger 2} a^2 a^\dagger a  + \frac{2}{45} (a^\dagger a)^3 \nonumber\\~&+&~ \frac{1}{60} a^{\dagger 3} a^3  + \frac{1}{90} a^{\dagger 2} a a^\dagger a^2 | s \rangle, 
\end{eqnarray}

\begin{eqnarray}\label{p1}
& P_1 & ~=~ \kappa^2\langle s | a^\dagger a | s\rangle - \kappa^4 \langle s | \frac{1}{3} (a^\dagger a)^2 + \frac{2}{3} a^{ \dagger 2} a^2 | s \rangle\nonumber \\
~&+&~ \kappa^6 \langle s | \frac{8}{45} a^{\dagger 2} a a^\dagger a^2 + \frac{4}{45} a^{\dagger 2}a^2 a^\dagger a + \frac{4}{45} a^\dagger a a^{\dagger 2}a^2 \nonumber\\&+& \frac{2}{45} (a^\dagger a)^3 + \frac{1}{10} a^{\dagger 3} a^3 | s \rangle
\end{eqnarray}

\begin{eqnarray}\label{p2}
P_2 ~&=&~ \frac{\kappa^4}{2} \langle s | a^{\dagger 2} a^2  |s \rangle - \kappa^6 \langle s | \frac{1}{4} a^{\dagger 3} a^3 + \frac{1}{6} a^{\dagger 2} a a^\dagger a^2  \nonumber\\&+& \frac{1}{24} a^{\dagger 2}a^2 a^\dagger a + \frac{1}{24} a^\dagger a a^{\dagger 2}a^2 | s \rangle,
\end{eqnarray} 
and,
\begin{equation}\label{p3}
P_3 ~=~ \frac{\kappa^6}{6} \langle s | a^{\dagger 3} a^3 | s \rangle.
\end{equation}

In deriving these formulae we have made no assumption about the commutator $[a^\dagger, a]$.

\section{Evolution of Coherent States and P Representation}
Equations~\eqref{po}-\eqref{p3} can be used directly to implement the criteria for acoherence to be discussed below for radiation fields that are characterized theoretically.  An alternative, complementary approach that is closer to procedures commonly used in quantum optics is also possible, as we now describe (also see Appendix.~\ref{AppExact}).  

For a coherent state $|\alpha \rangle$ of the field satisfying $a|\alpha \rangle=\alpha|\alpha\rangle$ with $[a,a^\dagger]=1$, and the detector in its ground state we have
\begin{eqnarray}
    &&U_I |\alpha\rangle\otimes| 0\rangle =e^{-i\sqrt{\gamma_0 \Delta t}(a^{\dagger}b+b^{\dagger} a)}|\alpha\rangle\otimes| 0\rangle \nonumber\\&=& e^{-i\sqrt{\gamma_0 \Delta t}(a^{\dagger}b+b^{\dagger} a)}e^{-|\alpha|^{2}/2}e^{\alpha a^{\dagger}}e^{i\sqrt{\gamma_0 \Delta t}(a^{\dagger}b+b^{\dagger} a)}\nonumber\\&\times&e^{-i\sqrt{\gamma_0 \Delta t}(a^{\dagger}b+b^{\dagger} a)}|0\rangle \otimes | 0\rangle=e^{-|\alpha|^{2}/2}e^{-i\sqrt{\gamma_0 \Delta t}(a^{\dagger}b+b^{\dagger} a)}\nonumber\\&\times&\bigg(\sum_{n}\frac{(\alpha a^{\dagger})^{n}}{n!}\bigg)e^{i\sqrt{\gamma_0 \Delta t}(a^{\dagger}b+b^{\dagger} a)}|0\rangle \otimes | 0\rangle\nonumber\\
    &=&e^{-|\alpha|^{2}/2}\bigg\{\sum_{n}\frac{[\alpha U_I a^{\dagger}U_I^{\dagger}]^{n}}{n!}\bigg\}|0\rangle  \otimes |0\rangle.
    \end{eqnarray}    
 From this, using $U_{I}a^{\dagger} U_I^{\dagger} = \cos(\sqrt{\gamma_0 \Delta t})a^{\dagger}-i\sin(\sqrt{\gamma_0 \Delta t})b^{\dagger}$, we find
    \begin{eqnarray}
   &&e^{-i\sqrt{\gamma_0 \Delta t}(a^{\dagger}b+b^{\dagger} a)}|\alpha\rangle \otimes | 0\rangle\nonumber\\   &=&e^{-|\alpha|^{2}/2}e^{\alpha\cos(\sqrt{\gamma_0 \Delta t})a^{\dagger}-i\alpha\sin(\sqrt{\gamma_0 \Delta t})b^{\dagger}}|0\rangle\otimes|0\rangle,\nonumber\\&=&|\alpha\cos(\sqrt{\gamma_0 \Delta t})\rangle \otimes |-i\alpha\sin(\sqrt{\gamma_0 \Delta t})\rangle.
\end{eqnarray}

A general density matrix can be expressed in the $P$ representation as~\cite{Sudharshan,Glauber,Mandel_Wolf_1995} 
\begin{equation}
   \rho = \int d^{2}\alpha P(\alpha)|\alpha\rangle\langle\alpha|.
\end{equation}
Thus the density of the joint system after a time $\Delta t$ is given by

\begin{eqnarray}
    &&U_{I}(\rho\otimes |0\rangle\langle 0|) U_{I}^{\dagger} = \int d^{2}\alpha P(\alpha)U_{I}|\alpha\rangle\langle\alpha|\otimes |0\rangle\langle 0|U_{I}^{\dagger}\nonumber\\&=&\int d^{2}\alpha P(\alpha)|\alpha \cos(\sqrt{\gamma_0 \Delta t})\rangle\langle\alpha \cos(\sqrt{\gamma_0 \Delta t})|\nonumber\\&\otimes& |-i\alpha \sin(\sqrt{\gamma_0 \Delta t})\rangle\langle -i\alpha\sin(\sqrt{\gamma_0 \Delta t})|.\nonumber\\
\end{eqnarray}

We obtain that the probability, $P_{n}$ is given by,
\begin{eqnarray}
   &&P_{n} =  \text{tr}_{F}\{\langle n|U_{I}(\rho\otimes |0\rangle\langle 0|) U_{I}^{\dagger}|n\rangle\}\nonumber\\&=&\frac{[\sin^2(\sqrt{\gamma_0\Delta t})]^{n}}{n!}\int d^{2}\alpha P(\alpha)|\alpha|^{2n}e^{-|\alpha|^2\sin^2(\sqrt{\gamma_0\Delta t})}.\nonumber\\
   \label{eqExact1}
\end{eqnarray}
After the approximate substitution $\sin^2(\sqrt{\gamma_0\Delta t})\rightarrow \gamma_0\Delta t$
this reproduces the result found in ~\cite{Mandel_Wolf_1995}, with a simpler derivation. The limit can also be arrived at using qubit detectors, as explored in Appendix.~\ref{qubitD}. The need for modification to the standard result from Ref.~\cite{Mandel_Wolf_1995} has already been discussed, see for example, Ref.~\cite{srinivas_photon_1981}. 

In Appendix.~\ref{AppExact}, we show how to recover the consequences of Eqns.\,\eqref{po}-\eqref{p3} from this standpoint.  The generalization of Eq.~\eqref{eqExact1} to finite detector efficiencies is also discussed in Appendix.~\ref{AppExact}.

\section{Global Counting Statistics\label{Gstat}}
Using Eq.~\eqref{eqExact1}, the average occupation that would be measured in the resonant harmonic detector is given by,
\begin{eqnarray}
    \bar{n} &=& \sum_{n=0}^{\infty} n P_{n} =\sum_n \frac{[\sin^2(\sqrt{\gamma_0\Delta t})]^{n}}{(n-1)!}\nonumber\\&\times&\int  d^{2}\beta P(\beta) |\beta|^{2n}e^{-|\beta|^2\sin^2(\sqrt{\gamma_0\Delta t})}\nonumber\\  &=&\sin^2(\sqrt{\gamma_0\Delta t})\int  d^{2}\beta |\beta|^{2} P(\beta) = \sin^2(\sqrt{\gamma_0\Delta t}) \langle a^\dagger a\rangle_{\rho}, \nonumber\\
\end{eqnarray}
and similarly
\begin{equation}
    \overline{n(n-1)} = \sum_{n=0}^{\infty} n(n-1) P_{n} = [\sin^2(\sqrt{\gamma_0\Delta t})]^2 \langle (a^\dagger)^2 a^2\rangle_{\rho}.
\end{equation}
Using this, one can write the variance as
\begin{equation}
    (\Delta n)^2 \approx\bar{n}+(\gamma_0\Delta t)^2 Q \langle n\rangle,\label{eqvar}
\end{equation}
where $Q$ is the Mandel's $Q$ parameter for the density matrix $\rho$, defined as the ratio,
\begin{eqnarray}
 Q=   \frac{\langle(\Delta \hat{N})^2\rangle_\rho-\langle \hat{N}\rangle_\rho}{\langle \hat{N}\rangle_\rho},
\end{eqnarray}
with $\hat{N} = a^{\dagger}a$. 
For a coherent state, which exhibits Poisson statistics for the counts, we have $Q\rightarrow 0$ and $(\Delta n)^2 = \bar{n}$.  
For a thermal state we have the $P$ representation
\begin{equation}
    P(\alpha) = \frac{1}{\pi n_{\text{th}}} e^{-|\alpha|^2/n_{\text{th}}}.
\end{equation}
leading to the super-Poissonian (variance $>$ mean) behavior
 \begin{equation}
     \bar{n} = \gamma_0\Delta t n_{\text{th}},~~Q=n_{\text{th}},~~(\Delta n)^2 = \bar{n}+(\gamma_0\Delta t)^2 n_{\text{th}}^2.
 \end{equation}

For a squeezed vacuum state
\begin{equation}
    |\psi_{sq}\rangle = \frac{1}{\sqrt{\cosh(r)}}\sum_{m=0}^{\infty}(-\tanh(r))^{m}\frac{\sqrt{2m!}}{2^m m!}|2m\rangle.
\end{equation}
we have
\begin{eqnarray}
    \langle n\rangle &=& \sinh(r)^2,~~\bar{n}=\gamma_0\Delta t\sinh(r)^2,\nonumber
\\ (\Delta n)^2 &=&\bar{n}+(\gamma_0\Delta t)^2 \cosh(2r)\sinh(r)^2, ~~Q= \cosh(2r),\nonumber\\
\end{eqnarray}
which is also super-Poissonian.  For equal intensity of the radiation field (same $\bar{n}$), the quantum noise induced by a highly squeezed vacuum state is approximately two times the noise induced by a thermal state.  

\section{Ratio Test\label{ratio-test}}
The Poissonian distribution induced by coherent states entails rigid relations between the probabilities for measuring small levels of excitation.  These lead to simple yet sharp quantitative signals for acoherence, as we now discuss.   
Applying  Eqn.~\eqref{eqExact1} to a coherent state we find
\begin{equation}\label{test1}
   R \, \equiv \, \frac{2P_{2}P_{0}}{P_{1}^2} = 1
\end{equation} 
and 
\begin{equation}\label{text2}
    R' \, \equiv \, \frac{3P_{3}P_{1}}{2P_{2}^2} =1 
\end{equation}
for coherent states. 


For Fock states (i.e., number eigenstates) the $P$ representation is awkward, but we can use Eqns.\,\eqref{po}-\eqref{p3}.  To lowest order, we find $P_{0}= 1$, $P_{1} = \gamma_0 \Delta t n $, $P_{2}=(\gamma_0 \Delta t)^2 n(n-1)/2$, $P_{3}=(\gamma_0 \Delta t)^3n(n-1)(n-2)/6$.  
\begin{eqnarray}
R ~&=&~ 1 - \frac{1}{n} \ , \nonumber \\
R^\prime ~&=&~ 1 - \frac{1}{n-1}
\end{eqnarray}
For small $n$ we are dealing with few quanta, so big deviations from classical behavior are to be expected (and our ratios can even bring in division by zero).  For large $n$ the deviations from $R = R^\prime =1$ approach zero. Nevertheless these are maximally sub-Poissonian states, with $Q= -1$.

For thermal states, using Eq.~\eqref{eqExact1}, we find 
\begin{equation}
 P_{0} = \frac{1}{1 + n_{\text{th}}[\sin^2(\sqrt{\gamma_0\Delta t})]}
\end{equation}
\begin{equation}P_{1} = \frac{[\sin^2(\sqrt{\gamma_0\Delta t})] n_{\text{th}}}{([\sin^2(\sqrt{\gamma_0\Delta t})] n_{\text{th}}+1)^2},   
\end{equation}
\begin{equation}
P_{2} = \frac{([\sin^2(\sqrt{\gamma_0\Delta t})] n_{\text{th}})^2}{([\sin^2(\sqrt{\gamma_0\Delta t})] n_{\text{th}}+1)^3}.    
\end{equation}
so that 
    \begin{equation}
 R= \frac{2P_{2}P_{0}}{P_{1}^2}=2.
\end{equation}
Remarkably, the ratio is independent of temperature. 


For squeezed vacuum states it is again convenient to use Eqns.\,\eqref{po}-\eqref{p3}.  To lowest order, we find 
\begin{equation}
 R= \frac{2P_{2}P_{0}}{P_{1}^2}=2+\coth ^2(r).
\end{equation}
For large squeezing, $\lim_{r\rightarrow\infty}\coth ^2(r)\rightarrow 1$ and so 
$ \frac{2P_{2}P_{0}}{P_{1}^2} \rightarrow 3$. 

We find that for coherent states and highly excited Fock states, our ratio $R$ probabilities is approximately one, while for thermal states it is approximately two, and for highly squeezed vacuum states it is approximately three. It also follows from observing that, using Eqns.\,\eqref{po}-\eqref{p3} to leading order,
\begin{equation}
    R\approx 1+Q/\langle n\rangle.\label{ratio}
\end{equation}
This shows that for large $\langle n\rangle$ sub-Poissonian states with $-1\leq Q<0$ have $R\sim 1$, making them hard to discriminate from coherent states. In Appendix~\ref{arbGaus}, we show how to evaluate the relevant count rates and the ratio $R$ for a thermal state of average quanta $n_{\text{th}}$ that is both displaced (in the phase space along the $x$ direction by $x_0$) and squeezed (by amplitude $r$ and phase $\phi$ relative to the direction of the displacement). To leading order, we find that,
\begin{widetext}
\begin{eqnarray}
    R&\approx& \frac{4 n_{\text{th}}^2-8 n_{\text{th}} x_{0}^2 \cos (\phi) \sinh (2 r)+8 (2 n_{\text{th}}+1) \left(x_{0}^2-1\right) \cosh (2 r)}{2 \left((2 n_{\text{th}}+1) \cosh (2 r)+x_{0}^2-1\right)^2}\nonumber\\&+&\frac{3 (2 n_{\text{th}}+1)^2 \cosh (4 r)+4 n_{\text{th}}-8 x_{0}^2 \cos (\phi) \sinh (r) \cosh (r)+2 x_{0}^4-8 x_{0}^2+5}{2 \left((2 n_{\text{th}}+1) \cosh (2 r)+x_{0}^2-1\right)^2}.
\end{eqnarray}
    
\end{widetext}

Here we see substantial deviation from $R=1$ for a generic Gaussian state of the radiation field.  Note that for a highly occupied field mode, $1\leq R\leq 3+\text{cosech}^{2}(r)$, with equality possible at both ends.

It can be verified that, to the leading order, the ratio $R$ equals the second-order coherence function for the radiation field.  The novelty here is that it is estimated here using only the lowest order probabilities in a resonant (click) detector. Many ways to access quantum properties of radiation fields have been proposed in experimental contexts such as in quantum optics where the detection capabilities are substantially more flexible. For example,  Mandel's $Q$ can be estimated straightforwardly for quantum light from global counting statistics, using good photodetectors, even at very low intensities.  Interferometric techniques also offer an alternate approach to estimating the second-order coherence function for quantum light~\cite{Mandel_Wolf_1995}. But the problem can be substantially more challenging if the radiation field in question is very weakly interacting, and when one expects to see only very few clicks within any realistic experimental window. This scenario is characteristic of the gravitational context and a number of other contexts briefly reviewed in Sec.~\ref{applications}. In the gravitational context, astrophysical scenarios such as binary black hole or binary neutron star merger events produce detectable strain amplitudes that only last at most a few seconds as seen by our detectors.  Here our approach appears most appropriate.  Here our approach appears most appropriate.   

While we primarily focus on probing the acoherence of radiation fields for large $\langle n\rangle$ (which is the relevant limit for gravitational radiation), it is evident from Eq.~\eqref{ratio} that our ratio test can also be used to test strict non-classicality in terms of sub-Poissonian statistics $(R<1)$, when the detector can register clicks at low intensities (eg., in quantum optics). A comparable ratio test for non-classicality using the width of the marginal distributions in phase-space has been discussed in Refs.~\cite{phasespace,PhaseSpace2} to probe the sub-Poissonian nature of optical fields. Tight inequalities involving lowest order click probabilities to probe non-classicality have also been discussed in a very recent work~\cite{Inequality}. 

\section{Possible Physical Applications\label{applications}}
Recent work ~\cite{tobar_detecting_2024} has made it plausible that resonant detection of gravitational waves, producing small levels of excitation, is a challenging but achievable goal.  In this context the issue of counting statistics is especially interesting, since -- as we have shown -- it can probe essentially quantum features of the radiation.  

It can be proved that linear coupling to deterministic (classical) sources produce coherent states of radiation fields, but stochastic or nonlinear couplings will generally bring in acoherence.  Quadratic coupling to a deterministic source, for example, brings in squeezed states.  We have demonstrated that acoherence can have quantitatively significant effects on the small count statistics even in cases where the underlying radiation fields have high intensity but are weakly coupled to the detector.   Perhaps ironically, ``bad'' detectors, with low count rates, reveal the effects most clearly, since they are especially sensitive to fluctuations.  Of course, it is possible to synthesize bad detectors from good ones, for example by sampling a slowly varying source in small time intervals.   Also, for long wavelength radiation, one can sample from an array of bad detectors.  

For the fundamental acoustic mode of the Weber bar as the resonant mass detector for gravitational waves treated quantum mechanically, the spontaneous emission rate $\gamma_0$ is given by (see Ref.~\cite{tobar_detecting_2024}),
\begin{equation}
   \gamma_0 = \frac{8GML^2\omega^4}{\pi^4 c^5},
\end{equation}
where $G$ is the Newton's constant, $M,L$ are respectively the mass and length of the Weber bar, $\omega$ is the resonant frequency, and $c$ is the speed of light. Using their values, the spontaneous emission rate is estimated to be rather very small $\sim 10^{-33}s^{-1}$ for a typical resonant bar detector for gravitational waves in the kHz range where LIGO operates~\cite{tobar_detecting_2024}. 
The deviation from Poissonian statistics has the form $\sim(\gamma_0\Delta t)^2 Q \langle n\rangle $, where $Q$ is the Mandel's parameter, and $\langle n\rangle $ is the number of quanta of the radiation field in the local mode volume. The corresponding energy flux density at frequency $\omega$ would be $\sim \text{(energy density)}\times c\sim \langle n\rangle \hbar\omega^4/c^2$.
   For a gravitational wave in the LIGO band, $\langle n\rangle \sim 10^{36}$~\cite{Dyson}. 
In the interaction between a LIGO band gravitational wave and a typical bar detector for the duration of $\Delta t$, we can have measures of acoherence $(\gamma_0\Delta t)^2 Q \langle n\rangle$.  Therefore although $\gamma_0\Delta t\ll 1$ for a typical bar detector, a large Mandel's $Q\sim O(\langle n\rangle$)    can make deviations from a pure coherent state of the field experimentally detectable by comparing the global statistics, or by using the ratio test. 

To make further estimates for $\Delta t$, note that gravitational waves produced by merger events are not typically monochromatic, but have a time-varying frequency $\omega \equiv\omega(t)$. Following Ref.~\cite{tobar_detecting_2024}, an upper bound for $\Delta t$ can be thought of as the time a gravitational wave spends in the resonance window of our detector, $\Delta t_{\text{max}}\approx t[\omega+\Delta\omega]-t[\omega-\Delta\omega]\approx 2\Delta\omega/(k\omega^{11/3})$ where $k = 48/5 [GM_c/(2c^3)]^{5/3}$. By requiring that the timescale and bandwidth are constrained to be within the full width at half maximum of the detector's response (sinc) function, one can estimate that the corresponding frequency window (or the detector bandwidth) should at least be $2\Delta \omega = 8/(\Delta t_{\text{max}})$. This results in the following upper bound for $\Delta t$ for chirping gravitational waves: $\Delta t_{\text{max}}\approx 2\sqrt{2/k}~\omega^{-11/6}$~\cite{tobar_detecting_2024}. Considering the event GW150914 having $M_c \sim 30 M_{\text{sol.}}$~\cite{PhysRevLett.116.061102}, we can estimate a $\Delta t_{\text{max}}\sim 5 \text{ms}$ at $\nu = 200\text{Hz}~(\omega = 2\pi\nu)$. Note that this very first gravitational wave signal observed by LIGO lasted approximately $200\text{ms}$, with a time-varying frequency between $35\text{Hz}-250\text{Hz}$.   Thus meaningful tests of acoherence appear to be accessible on realistic timescales using the sorts of resonant mass detectors proposed in Ref.~\cite{tobar_detecting_2024} for gravitational radiation in the LIGO band. 

Gravitational radiation produced by neutron star---neutron star merger events such as GW170817~\cite{abbott2017gw170817} (which was suggested as a good candidate event for detecting a single graviton in Ref.~\cite{tobar_detecting_2024}) can also
   be good candidate events to see gravitational wave acoherence. Binary neutron star merger events appear to have slowly varying chirp frequencies and the gravitational wave can be seen in LIGO for up to a duration of a few seconds. Since they have slowly varying chirps, the gravitational waves can be approximated as monochromatic waves within small time windows $\Delta t$.  Considering $M_c\sim 1.19M_{\text{sol.}}$ we can estimate
$\Delta t_{\text{max}}\sim 4\text{ms}$ at kHz frequencies, and 
 $\sim 70\text{ms}$ around $\nu=200$Hz  . Thus it is highly suggestive that several detections with a $\Delta t\sim 1 \text{ms}$
    can be made using independent resonant mass detectors (or with rapid resetting of a single detector) to map out the quantum statistics of absorbed gravitons using the methods proposed here. 

Since the ratio test is practically significant at low count rates, it can be prone to additional experimental errors, in addition to the pathologies (of it being a ratio) we already pointed out. In these contexts, the inference problem can also be mapped into an interesting hypothesis-testing problem, which can help assign a certain fidelity to the degree of acoherence estimated using the ratio test. 

The departures from coherent states in gravitational waves might also be observable using interferometric detectors~\cite{ParikhPRL,ParikhPRD,ParikhEPJD}.  The class of quantum states that are expected to induce measurable quantum noise in interferometric detectors (highly squeezed vacuum states and thermal states~\cite{ParikhPRL,ParikhPRD,ParikhEPJD}) are the same ones that we predict would show quantifiable departures from coherent states of gravitational radiation in resonant mass detectors. When the gravitational radiation is in one of these quantum states, the quantum noise is super-Poissonian, making it potentially observable in both interferometric and resonant detectors. Of course, use of independent detection schemes would help to decrease both statistical and systematic uncertainties.

In the context of gravitational radiation, strongly nonlinear sources arise in the final stages of black hole mergers and in subsequent ring-downs.  Here the driving field is robust, so deviation from Poisson statistics would implicate quantum theory.  (The feasibility of detecting quantum effects in gravity is much debated; see {\it e. g}. ~\cite{rothman_can_2006,Dyson,Boughn_2006,CarneyPRD,carney_comments_2024,shenderov_stimulated_2024}, in particular the difficulty in seeing sub-Poissonian statistics of gravitational radiation from global statistics was pointed out recently in~\cite{carney_comments_2024,CarneyPRD}.)  Note that ring-down through quasinormal modes is expected to depend on only a few parameters of the final state, so that one can aspire to compare and combine statistics from distinct sources, separated in space and time.  Neutron star mergers, and possibly neutron star - black hole mergers, will plausibly involve complex, effectively stochastic processes as the neutron stars deform and shatter, again bringing in acoherence. 

Outside these fundamental but challenging applications to gravitational radiation, the issue of acoherence arises in many other contexts, notably including the quantum description of single mode lasers, where it impacts the feasibility of continuous variable quantum teleportation~\cite{rudolph_requirement_2001,wiseman_comment_2003}.
It can also be relevant for quantum acoustics, where substantial progress has been made in the recent years in creating and characterizing acoherent states of mechanical modes~\cite{chu_creation_2018}.   
Several recent experiments generate non-classical radiation fields in other settings, notably including spontaneous emission of matter waves from ultracold atomic assemblies~\cite{krinner_spontaneous_2018}. 

\noindent\textit{Acknowledgments.---}
FW is supported by the U.S. Department of Energy under grant Contract Number DE-SC0012567 and by the Swedish Research Council under Contract No. 335-2014-7424. SKM is supported in part by the Swedish Research Council under Contract No. 335-2014-7424 and in part by the Wallenberg Initiative on Networks and Quantum Information (WINQ).  We thank Maulik Parikh and Igor Pikovski for stimulating conversations around these subjects. 
\appendix

\begin{widetext}
 \section{Identifying $\gamma_0$ as the spontaneous emission rate of the detector\label{graviton}}
A generic interaction Hamiltonian for resonant interactions between a continuous radiation field and a resonant detector in the interaction picture to leading order has the following form,
\begin{equation}
    H_{I}(t) = \hbar g\sum_\nu [a^\dagger b e^{i(\nu-\omega)t}+b^{\dagger}ae^{-i(\nu-\omega)t}],
\end{equation}
where $g$ is the coupling that captures the details of the interaction and $\sum_{\nu}$ represents summation over field modes near resonance to the detector. As an example, for the resonant bar detector for gravitational waves considered in Ref.~\cite{tobar_detecting_2024}, $g = \sqrt{\frac{8GML^2\omega^5}{\pi^3 c^5}}.$ Our objective here is to map the resultant dynamics to a simplified single mode approximation, satisfying,
\begin{equation}
    H_{I}t = \hbar\sqrt{\gamma_0 t}(a^{\dagger} b+b^{\dagger} a),
\end{equation}
with $\sqrt{\gamma_0}$ as the effective coupling, where $\gamma_0$ is the spontaneous emission rate of the detector.

 In order to do so, we consider the resonant detector initially in the ground state and follow the standard approach to derive the Fermi-Golden rule transition rates. Considering the resonant detector in the first excited state, and the field is in the vacuum state, their joint state after time $t$ to the leading order is given by,
\begin{equation}
|\psi'(t)\rangle=\sum_\nu-ig e^{i(\nu-\omega)t/2}\frac{2\sin[(\omega-\nu)t/2]}{(\omega-\nu)} a_{\nu}^{\dagger}|0\rangle_{F}b|1\rangle_{\text{D}}.
\end{equation}
This yields the following probability of observing a single quantum of the radiation field spontaneously emitted by the detector, given by, 
\begin{equation}
   P_{\text{spont.}}\approx \sum_\nu 4 g^2\frac{\sin^2[(\nu-\omega)t/2]}{(\nu-\omega)^2}=4g^2D(\omega)\int_{\omega-\delta/2}^{\omega+\delta/2}d\nu \frac{\sin^2[(\nu-\omega)t/2]}{(\nu-\omega)^2}, 
\end{equation}
where we have converted the summation to an integral using the density of field modes $D(\omega)$. For $t\delta >> 1$, we have~\cite{loudon_quantum_2000},
\begin{equation}
    \int_{\omega-\delta/2}^{\omega+\delta/2}d\nu \frac{\sin^2[(\nu-\omega)t/2]}{(\nu-\omega)^2}\approx \frac{1}{2}\pi t. 
\end{equation}
Hence we obtain the probability of spontaneous emission as,
\begin{equation}
     P_{\text{spont.}}\approx 2g^2D(\omega) \pi t.
\end{equation}
The density of states $D(\omega)$ of plane waves per volume $V$  for a given polarization is,
\begin{equation}
   \sum_{\Omega} = \int d\Omega D(\Omega) = \int d\Omega \frac{V\Omega^2}{2\pi^{2}c^3}.
\end{equation} 
We can take the volume to be the reduced volume of a quanta of the field, $V=(c/\omega)^3$, which yields, $D(\omega)=\frac{1}{2\pi^{2}\omega}.$ The probability of registering a click becomes,
\begin{equation}
   P_{\text{spont.}}\approx 2g^2D(\omega) \pi t = \frac{g^2 t}{\pi\omega} = \gamma_0 t,\label{eqSE}
\end{equation}
where $\gamma_0$ is the spontaneous emission rate. Similarly, the state of the field and the detector when the detector is initially in the ground state and the field is in an arbitrary quantum state $|\psi_F(0)\rangle$ is given by,
\begin{equation}
|\psi(t)\rangle=\sum_\nu-ig e^{-i(\nu-\omega)t/2}\frac{2\sin[(\nu-\omega)t/2]}{(\nu-\omega)} a_{\nu}|\psi_F(0)\rangle b^{\dagger}|0\rangle_{\text{D}}.
\end{equation}
The probability of registering a click in the detector is given by,
\begin{equation}
   P_{1}\approx \sum_\nu 4 g^2\frac{\sin^2[(\nu-\omega)t/2]}{(\nu-\omega)^2} \langle a_{\nu}^{\dagger}a_{\nu}\rangle_F=4g^2D(\omega)\langle a_{\omega}^{\dagger}a_{\omega}\rangle_F\int_{\omega-\delta/2}^{\omega+\delta/2}d\nu \frac{\sin^2[(\nu-\omega)t/2]}{(\nu-\omega)^2}.  
\end{equation}
Again taking the limit $t\delta >> 1$, we obtain,

\begin{equation}
   P_{1}\approx 2g^2D(\omega)\langle a_{\omega}^{\dagger}a_{\omega}\rangle_F \pi t = \frac{g^2 t}{\pi\omega} \langle a_{\omega}^{\dagger}a_{\omega}\rangle_F =\gamma_0 \langle a_{\omega}^{\dagger}a_{\omega}\rangle t,
\end{equation}
where we have identified $\gamma_0$ as the spontaneous emission rate from Eq.~\eqref{eqSE}. We can now map the dynamics to an effective measurement model, noticing that an interaction Hamiltonian satisfying,
\begin{equation}
    H_{I}t = \hbar\sqrt{\gamma_0 t}(a^{\dagger} b+b^{\dagger} a),
\end{equation}
reproduces the probabilities derived above in the single mode approximation for the field. This allows us to identify the parameter $\gamma_0$ in our effective measurement model discussed in the main text, as the spontaneous emission rate, an intrinsic property of our detector.

\section{Exact solution to the unitary dynamics on a coherent state of the field\label{AppExact}}
We are interested in computing the general result for the field modes in a coherent state,
\begin{eqnarray}
    e^{-i\sqrt{\gamma_0 \Delta t}(a^{\dagger}b+b^{\dagger} a)}|\alpha\rangle| 0\rangle &=& e^{-i\sqrt{\gamma_0 \Delta t}(a^{\dagger}b+b^{\dagger} a)}e^{-|\alpha|^{2}/2}e^{\alpha a^{\dagger}}e^{i\sqrt{\gamma_0 \Delta t}(a^{\dagger}b+b^{\dagger} a)}e^{-i\sqrt{\gamma_0 \Delta t}(a^{\dagger}b+b^{\dagger} a)}|0\rangle| 0\rangle\nonumber\\
    &=&e^{-|\alpha|^{2}/2}e^{-i\sqrt{\gamma_0 \Delta t}(a^{\dagger}b+b^{\dagger} a)}\bigg(\sum_{n}\frac{(\alpha a^{\dagger})^{n}}{n!}\bigg)e^{i\sqrt{\gamma_0 \Delta t}(a^{\dagger}b+b^{\dagger} a)}|0\rangle| 0\rangle\nonumber\\
    &=&e^{-|\alpha|^{2}/2}\bigg\{\sum_{n}\frac{[\alpha e^{-i\sqrt{\gamma_0 \Delta t}(a^{\dagger}b+b^{\dagger} a)} a^{\dagger}e^{i\sqrt{\gamma_0 \Delta t}(a^{\dagger}b+b^{\dagger} a)}]^{n}}{n!}\bigg\}|0\rangle| 0\rangle
    \end{eqnarray}
    Above, we have used the identity that $U_{I}(a^{\dagger})^n U_I^{\dagger}|0\rangle|0\rangle = (U_{I}a^{\dagger} U_I^{\dagger})^n |0\rangle|0\rangle$. We can therefore simplify, using, $U_{I}a^{\dagger} U_I^{\dagger} = \cos(\sqrt{\gamma_0 \Delta t})a^{\dagger}-i\sin(\sqrt{\gamma_0 \Delta t})b^{\dagger}$ to obtain,
    \begin{eqnarray}
   e^{-i\sqrt{\gamma_0 \Delta t}(a^{\dagger}b+b^{\dagger} a)}|\alpha\rangle| 0\rangle   &=&e^{-|\alpha|^{2}/2}e^{\alpha\cos(\sqrt{\gamma_0 \Delta t})a^{\dagger}-i\alpha\sin(\sqrt{\gamma_0 \Delta t})b^{\dagger}}|0\rangle|0\rangle,\nonumber\\&=&|\alpha\cos(\sqrt{\gamma_0 \Delta t})\rangle|-i\alpha\sin(\sqrt{\gamma_0 \Delta t})\rangle.
\end{eqnarray}
We can now use the result for a coherent state of the field obtained above to compute the time evolution of an arbitrary, but unknown quantum state of the field $\rho$, using the P representation for the field. In the P representation, we may write,
\begin{equation}
   \rho = \int d^{2}\alpha P(\alpha)|\alpha\rangle\langle\alpha|.
\end{equation}
The state of the joint system after a time $\Delta t$ is given by,
\begin{eqnarray}
    &&U_{I}(\rho\otimes |0\rangle\langle 0|) U_{I}^{\dagger} = \int d^{2}\alpha P(\alpha)U_{I}|\alpha\rangle\langle\alpha|\otimes |0\rangle\langle 0|U_{I}^{\dagger}\nonumber\\&\approx&\int d^{2}\alpha P(\alpha)|\alpha \cos(\sqrt{\gamma_0 \Delta t})\rangle\langle\alpha \cos(\sqrt{\gamma_0 \Delta t})|\otimes |-i\alpha \sin(\sqrt{\gamma_0 \Delta t})\rangle\langle -i\alpha\sin(\sqrt{\gamma_0 \Delta t})|.\nonumber\\
\end{eqnarray}
We obtain that the probability, $P_{n}$ is given by,
\begin{equation}
   P_{n} =  \text{tr}_{F}\{\langle n|U_{I}(\rho\otimes |0\rangle\langle 0|) U_{I}^{\dagger}|n\rangle\}=\frac{[\sin^2(\sqrt{\gamma_0\Delta t})]^{n}}{n!}\int d^{2}\alpha P(\alpha)|\alpha|^{2n}e^{-|\alpha|^2\sin^2(\sqrt{\gamma_0\Delta t})}.\label{eqExact}
\end{equation}
Note that the result is naturally in the normally ordered form as it is obtained using the P representation, and agrees to its standard form given in Ref.~\cite{Mandel_Wolf_1995} when we approximate $\sin^2(\sqrt{\gamma_0\Delta t})\approx \gamma_0\Delta t$.

Generalization to finite detector efficiencies is also straightforward by considering an additional beam-splitter coupled to the detector mode with transmission and reflection amplitudes $\sqrt{\eta}$ and $\sqrt{1-\eta}$ respectively with one of the beam-splitter inputs being the signal transmitted with probability $\eta$, the detector efficiency, and other port in the vacuum that gets populated with the signal probability $1-\eta$ which will be unobserved. Following a similar approach to above, it can be shown that the observable click probabilities at finite detector efficiency $\eta$ gets modified to,
    \begin{equation}
        P_n = \frac{[\eta \sin^2(\sqrt{\gamma_0\Delta t})]^{n}}{n!}\int d^{2}\alpha P(\alpha)|\alpha|^{2n}e^{-\eta|\alpha|^2\sin^2(\sqrt{\gamma_0\Delta t})}.
    \end{equation}

For a coherent state of the field, we can also series expand the exact probabilities computed above in Eq.~\eqref{eqExact} to the third order to find that (for $\kappa =\sqrt{\gamma_0\Delta t}$),
\begin{eqnarray}
   P_{0} &=& \frac{1}{90} |\alpha|^2 \kappa^2 \left(-\left(15 \left(|\alpha|^2+2\right) |\alpha|^2+4\right) \kappa^4+15 \left(3 |\alpha|^2+2\right) \kappa^2-90\right)+1,\nonumber\\
   P_{1} &=& \frac{1}{90} |\alpha|^2 \kappa^2 \left(-30 \left(3 |\alpha|^2+1\right) \kappa^2+\left(45 |\alpha|^4+60 |\alpha|^2+4\right) \kappa^4+90\right),\nonumber\\
   P_{2}&=& \frac{1}{6} |\alpha|^4 \kappa^4 \left(3-\left(3 |\alpha|^2+2\right) \kappa^2\right),\label{eqcoh1}
\end{eqnarray}
and,
\begin{equation}
    P_{3} = \frac{|\alpha|^6 \kappa^6}{6}.\label{eqcoh2}
\end{equation}
Here we show that they agree to the probabilities computed in the main text using their normally ordered form,\begin{eqnarray}
    &P_0& ~=~ \langle s|\bigg\{ 1 - \kappa^2 a^{\dagger}a + \kappa^4 \bigg[\frac{1}{6} a^{\dagger 2}a^2 + \frac{1}{3} (a^{\dagger}a+a^{\dagger 2}a^2)\bigg]  \\
   ~&-&~ \kappa^6 \bigg[  \frac{17}{180} (2a^{\dagger 2}a^2+a^{\dagger 3}a^3)  + \frac{2}{45} (a^{\dagger}a+3a^{\dagger 2}a^2+a^{\dagger 3}a^3) + \frac{1}{60} (a^{\dagger 3}a^3)  + \frac{1}{90} (a^{\dagger 2}a^2+a^{\dagger 3}a^3) \bigg]\bigg\}|s\rangle ,\nonumber 
\end{eqnarray}
\begin{eqnarray}
& P_1 & = \langle s|\bigg\{\kappa^2 a^{\dagger}a - \kappa^4 \bigg[\frac{1}{3} (a^{\dagger}a+a^{\dagger 2}a^2) + \frac{2}{3} (a^{\dagger 2}a^2) \bigg]\bigg\}|s\rangle \\
~&+&~ \kappa^6 \bigg[\frac{8}{45} (a^{\dagger 2}a^2+a^{\dagger 3}a^3) + \frac{8}{45} (2a^{\dagger 2}a^2+a^{\dagger 3}a^3) + \frac{2}{45} (a^{\dagger}a+3a^{\dagger 2}a^2+a^{\dagger 3}a^3) + \frac{1}{10} (a^{\dagger 3}a^3) \bigg]\bigg\}|s\rangle, \nonumber
\end{eqnarray}
\begin{equation}
P_2 ~=~ \langle s|\bigg\{\frac{\kappa^4}{2} a^{\dagger 2}a^2 - \kappa^6 \bigg[ \frac{1}{4} (a^{\dagger 3}a^3) + \frac{1}{6} (a^{\dagger 2}a^2+a^{\dagger 3}a^3)  + \frac{1}{12} (2a^{\dagger 2}a^2+a^{\dagger 3}a^3)\bigg]\bigg\}|s\rangle.
\end{equation} 
\begin{equation}
P_3 ~=~ \langle s|\bigg\{\frac{\kappa^6}{6} a^{\dagger 3}a^3\bigg\}|s\rangle.
\end{equation}
For a coherent state $|s\rangle = |\alpha\rangle,$ we can write,
\begin{eqnarray}
    &P_0& ~=~ 1 - \kappa^2 |\alpha|^2 + \kappa^4 \bigg[\frac{1}{6} |\alpha|^4 + \frac{1}{3} (|\alpha|^2+|\alpha|^4)\bigg]  \\
   ~&-&~ \kappa^6 \bigg[ \frac{17}{180} (2|\alpha|^4+|\alpha|^6) + \frac{2}{45} (|\alpha|^2+3|\alpha|^4+|\alpha|^6) + \frac{1}{60} (|\alpha|^6)  + \frac{1}{90} (|\alpha|^4+|\alpha|^6) \bigg] \nonumber 
\end{eqnarray}
\begin{eqnarray}
& P_1 & = \kappa^2|\alpha|^2 - \kappa^4 \bigg[\frac{1}{3} (|\alpha|^2+|\alpha|^4) + \frac{2}{3} (|\alpha|^4) \bigg] \\
~&+&~ \kappa^6 \bigg[\frac{8}{45} (|\alpha|^4+|\alpha|^6) + \frac{8}{45} (2|\alpha|^4+|\alpha|^6) + \frac{2}{45} (|\alpha|^2+3|\alpha|^4+|\alpha|^6) + \frac{1}{10} (|\alpha|^6) \bigg] \nonumber
\end{eqnarray}
\begin{equation}
P_2 ~=~ \frac{\kappa^4}{2} |\alpha|^4 - \kappa^6 \bigg[ \frac{1}{4} (|\alpha|^6) + \frac{1}{6} (|\alpha|^4+|\alpha|^6)  + \frac{1}{12} (2|\alpha|^4+|\alpha|^6)\bigg].
\end{equation} 
\begin{equation}
P_3 ~=~ \frac{\kappa^6}{6} |\alpha|^6.
\end{equation}
Note that upon simplification, these are identical to Eqs.~\eqref{eqcoh1}\&\eqref{eqcoh2}.
 \subsection{A useful approximation}
For a general Gaussian state of the field, however, the P function is often not easy to compute. In this case, an approximate result for the statistics is also useful. To this end, we note that the Baker-Campbell-Hausdorff (BCH) expansion can be used. The time evolution operator can now be approximated by the BCH expansion,
\begin{eqnarray}
    \exp(-iH\Delta t/\hbar) &\approx& \exp{(-i\sqrt{\gamma_0 \Delta t}b^{\dagger} a)}\exp{(-i\sqrt{\gamma_0 \Delta t}a^{\dagger} b)}\exp{(\gamma_0 \Delta t[b^{\dagger} a,a^{\dagger} b]/2)} \nonumber\\&=& \exp{(-i\sqrt{\gamma_0 \Delta t}b^{\dagger} a)}\exp{(-i\sqrt{\gamma_0 \Delta t}a^{\dagger} b)}\exp{(\gamma_0 \Delta t(b^\dagger b (1+a^{\dagger}a)-(a^\dagger a (1+b^{\dagger}b) )/2)}.\nonumber\\ 
\end{eqnarray}
The approximation is that we have neglected higher order non-commuting terms from the BCH expansion. Considering the $b$ mode initialized in the vacuum state, the above approximation is valid for valid for $\gamma_0 \Delta t \langle a^{\dagger}a\rangle \ll 1$. In this case, we can simplify,
\begin{eqnarray}
    \exp(-iH\Delta t/\hbar)   |\psi\rangle |0\rangle&\approx& \exp{(-i\sqrt{\gamma_0 \Delta t}b^{\dagger} a)}\exp{(-i\sqrt{\gamma_0 \Delta t}a^{\dagger} b)}\exp{(-\gamma_0 \Delta t a^\dagger a/2)}|\psi\rangle|0\rangle\nonumber\\
&\approx&\exp{(-i\sqrt{\gamma_0 \Delta t}b^{\dagger} a)}\exp{(-\gamma_0 \Delta t a^\dagger a/2)}|\psi\rangle|0\rangle\nonumber\\
&\approx&\sum_{n=0}^{\infty}\frac{(-i\sqrt{\gamma_0 \Delta t})^n}{\sqrt{n!}}|n\rangle a^n \exp{(-\gamma_0 \Delta t a^\dagger a/2)}|\psi\rangle.
\end{eqnarray}
From this, we can estimate the measurement operators for observing $n-$th excitation, given by,
\begin{equation}
    \hat{M}_n = \langle n|\exp(-iH\Delta t/\hbar)   |0\rangle \approx \frac{(-i\sqrt{\gamma_0 \Delta t})^n}{\sqrt{n!}} a^n \exp{(-\gamma_0 \Delta t a^\dagger a/2)}.
 \end{equation}
These measurement operators can also be used to describe continuous quantum measurements and generate quantum trajectories of resonant fluorescence from harmonically trapped quantum particles, in the spirit of Refs.~\cite{jordan_anatomy_2016,lewalle_measuring_2020}.  The probability of measuring $n-$th excitation is given by,
\begin{equation}
    P_{n} =\langle\psi|\hat{M}_n^{\dagger}\hat{M}_n|\psi\rangle \approx  \frac{(\gamma_0 \Delta t)^n}{n!}\langle\psi|\exp{(-\gamma_0 \Delta t a^\dagger a/2)}(a^{\dagger})^n a^n \exp{(-\gamma_0 \Delta t a^\dagger a/2)}|\psi\rangle\label{ArbApprox}
\end{equation}
For a coherent state, we can use,
\begin{equation}
    \exp{(-\gamma_0 \Delta t a^\dagger a/2)}|\alpha\rangle = e^{-\frac{|\alpha|^2}{2}(1-e^{-\gamma_0 \Delta t})}|\alpha e^{-\gamma_0 \Delta t/2}\rangle,\label{ProbE}
\end{equation}
and write the above probability for an initial coherent state as,
\begin{eqnarray}
    P_{n} &\approx& \frac{(\gamma_0 \Delta t)^n}{n!} \langle\alpha|\exp{(-\gamma_0 \Delta t a^\dagger a/2)}(a^{\dagger})^n a^n \exp{(-\gamma_0 \Delta t a^\dagger a)/2}|\alpha\rangle \nonumber\\&=& \frac{(\gamma_0 \Delta te^{-\gamma_0 \Delta t})^n}{n!} |\alpha|^{2n}e^{-|\alpha|^2(1-e^{-\gamma_0 \Delta t})}.
\end{eqnarray}
Note that the probability is approximately normalized. We have,
\begin{eqnarray}
    \sum_n P_{n} &=& \sum_n \frac{(\gamma_0 \Delta t e^{-\gamma_0 \Delta t})^n}{n!} |\alpha|^{2n}e^{-|\alpha|^2(1-e^{-\gamma_0 \Delta t})}=e^{-|\alpha|^2(1-e^{-\gamma_0 \Delta t})} \sum_n \frac{(\gamma_0 \Delta t e^{-\gamma_0 \Delta t})^n}{n!} |\alpha|^{2n}\nonumber\\&=&e^{-|\alpha|^2(1-e^{-\gamma_0 \Delta t})}e^{|\alpha|^2\gamma_0 \Delta t e^{-\gamma_0 \Delta t}} =e^{-|\alpha|^2(1-e^{-\gamma_0 \Delta t}-\gamma_0 \Delta t e^{-\gamma_0 \Delta t})}\nonumber\\&\approx& e^{-|\alpha|^2(1-1+\gamma_0 \Delta t-\gamma_0 \Delta t )}=1.
\end{eqnarray}
For an arbitrary density matrix, we can write,
\begin{eqnarray}
       P_{n} &=& \text{tr}\{\rho\hat{M}_n^{\dagger}\hat{M}_n\} =\int  d^{2}\beta P(\beta)\langle\beta|\exp{(-\gamma_0 \Delta t a^\dagger a/2)}(a^{\dagger})^n\frac{(\gamma_0 \Delta t e^{-\gamma_0 \Delta t})^n}{n!} a^n \exp{(-\gamma_0 \Delta t a^\dagger a/2)}|\beta\rangle\nonumber\\
       &=&\frac{(\gamma_0 \Delta t e^{-\gamma_0 \Delta t})^n}{n!}\int  d^{2}\beta P(\beta) |\beta|^{2n}e^{-|\beta|^2(1-e^{-\gamma_0 \Delta t})}.\label{pS}
\end{eqnarray}
In arriving at the above result, we have used the diagonal P representation for the density matrix in the basis of coherent states~\cite{Sudharshan,Glauber},
\begin{equation}
    \rho = \int d^{2}\beta P(\beta)|\beta\rangle\langle \beta|.
\end{equation}
Note that, owing to the approximation we made, our result in Eq.~\eqref{pS} varies slightly from Ref.~\cite{Mandel_Wolf_1995} or Eq.~\eqref{eqExact}. However, starting from Eq.~\eqref{pS} the standard result from Ref.~\cite{Mandel_Wolf_1995} is recovered by taking the limit, $e^{-\gamma_0\Delta t}\approx 1-\gamma_0\Delta t$, and $\gamma_0\Delta t e^{-\gamma_0\Delta t}\approx\gamma_0\Delta t,$ where we can approximate,
\begin{equation}
    P_{n}\approx \frac{(\gamma_0 \Delta t)^n}{n!}\int  d^{2}\beta P(\beta) |\beta|^{2n}e^{-|\beta|^2\gamma_0 \Delta t}.\label{pM}
\end{equation}
Returning to the approximate result Eq.~\eqref{ArbApprox} in the number representation, we note that Eq.~\eqref{ArbApprox} can be useful to estimate the counting statistics and the probabilities for a generic Gaussian state which can have a P-representation that is non-trivial. We discuss this in Appendix.~\ref{arbGaus}.


\section{Taking the limit of a qubit detector and series of clicks\label{qubitD}}
Here we consider an alternate derivation of the global statistics using a series of qubits functioning as detectors, for pedagogical reasons. To this end we approximate the time evolution operator by series,
\begin{eqnarray}
    \exp(-iH\Delta t/\hbar) &=& 1-i\sqrt{\gamma_0 \Delta t}(a^{\dagger} b+b^{\dagger} a)-\frac{\gamma_0 \Delta t}{2}[(a^{\dagger} b)^2+(b^{\dagger} a)^2+a^{\dagger } b b^{\dagger} a+b^{\dagger} aa^{\dagger} b]+...\nonumber\\
    &=&1-i\sqrt{\gamma_0 \Delta t}(a^{\dagger} b+b^{\dagger} a)-\frac{\gamma_0 \Delta t}{2} [(a^{\dagger} b)^2+(b^{\dagger} a)^2+a^{\dagger }  a (1\pm b^\dagger b)+b^{\dagger}b (1+a^{\dagger}a)]+...\nonumber\\
\end{eqnarray}
Above we have used $bb^{\dagger}=1\pm b^\dagger b$ depending on the commutation relations. We may now consider the initial state, $|\psi\rangle |0\rangle$, where the $b$ mode is initialized in vacuum. Here we show two alternate but simple approaches to get to Eq.~\eqref{eqlast}. 
The state evolves as,
\begin{eqnarray}
    &&1-i\sqrt{\gamma_0 \Delta t}(a^{\dagger} b+b^{\dagger} a)-\frac{\gamma_0 \Delta t}{2} [(a^{\dagger} b)^2+(b^{\dagger} a)^2+a^{\dagger }  a (1\pm b^\dagger b)+b^{\dagger}b (1+a^{\dagger}a)]+... |\psi\rangle|0\rangle \nonumber\\
    &\rightarrow&-i\sqrt{\gamma_0 \Delta t} a|\psi\rangle|1\rangle+ \bigg(1-\frac{\gamma_0 \Delta t}{2}a^{\dagger }  a \bigg)|\psi\rangle |0\rangle -\frac{\gamma_0 \Delta t}{2} a^2|\psi\rangle (b^{\dagger})^2|0\rangle.
\end{eqnarray}
We assume that we have a qubit detector so that $(b^{\dagger})^2 = 0$. First, note that in a single time step, the probability of exciting a qubit when the field is in a coherent state $|\psi\rangle = |\alpha\rangle$ is given by $P_1 = \epsilon |\alpha|^2,$ where $\epsilon = \gamma_0 \Delta t$, and the probability of not exciting the qubit is $P_0 =1- \epsilon |\alpha|^2$. Now if we approximate each detector events as independent (as for small $\epsilon$, a coherent state of the field remains the same to leading order) with probabilities $P_0$ and $1-P_0$, we arrive at Eq.~\eqref{EqB} from which Eq.~\eqref{eqlast} follows. 

Alternatively, for a series of detectors initialized in state $|0\rangle$ interacting sequentially with the radiation field (each for a brief duration $\Delta t$ only), we can write,
\begin{eqnarray}
|\psi\rangle |0\rangle &\rightarrow& (1-\frac{\gamma_0\Delta t}{2} a^{\dagger}a )|\psi\rangle |0\rangle-i\sqrt{\gamma_0 \Delta t}a |\psi\rangle |1\rangle]|0\rangle\nonumber\\&\rightarrow&(1-\frac{\gamma_0\Delta t}{2} a^{\dagger}a )[(1-\frac{\gamma_0\Delta t}{2} a^{\dagger}a )|\psi\rangle |0\rangle-i\sqrt{\gamma_0 \Delta t}a |\psi\rangle |1\rangle]|0\rangle-i\sqrt{\gamma_0 \Delta t}a[(1-\frac{\gamma_0\Delta t}{2} a^{\dagger}a )|\psi\rangle |0\rangle\nonumber\\&-&i\sqrt{\gamma_0 \Delta t}a |\psi\rangle |1\rangle]|1\rangle\nonumber\\&\rightarrow&...
\end{eqnarray}
In  a compact form, we may write the global state before measuring the qubits,
\begin{eqnarray}
|\psi\rangle |0\rangle &\rightarrow& \hat{M}_0|\psi\rangle |0\rangle+\hat{M}_1 |\psi\rangle |1\rangle\rightarrow(\hat{M}_0 |0\rangle+\hat{M}_1 |1\rangle)^{\otimes 2}|\psi\rangle \rightarrow ... 
\nonumber\\&\rightarrow&(\hat{M}_0 |0\rangle+\hat{M}_1 |1\rangle)^{\otimes N}|\psi\rangle\nonumber\\ &\approx& \sum_{j=0}^{N}\frac{\sqrt{N!}}{\sqrt{(N-j)!j!}}\{(\hat{M}_0)^{N-j}(\hat{M}_{1} )^{j}|\psi\rangle|\{0\}^{\otimes (N-j)}\{1\}^{\otimes j}\rangle.\nonumber\\
\end{eqnarray}
We have defined $\hat{M}_0 = (1-\frac{\gamma_0\Delta t}{2} a^{\dagger}a )$ and $\hat{M}_1 = -i\sqrt{\gamma_0 \Delta t}a$ and have neglected their commutator to the leading order. 
Here the state $|\{0\}^{\otimes (N-j)}\{1\}^{\otimes j}\rangle$ represents the normalized symmetrized state of $j$ qubit detectors excited among a total of $N$ detectors. We can simplify this result for a coherent state, $|\psi\rangle = |\alpha\rangle,$
\begin{equation}
    |\psi(j,N)\rangle =\sum_{j=0}^{N}\frac{\sqrt{N!}}{\sqrt{(N-j)!j!}}(-i\sqrt{\epsilon} \alpha )^{j}(\hat{M}_0)^{N-j}|\alpha\rangle|\{0\}^{\otimes (N-j)}\{1\}^{\otimes j}\rangle. 
\end{equation}
The probability of observing $j$ ticks in $N$ steps is same as the probability of measuring the symmetrized state $|\{0\}^{\otimes (N-j)}\{1\}^{\otimes j}\rangle$, given by,
\begin{equation}
    p(j,N) = \frac{N!}{(N-j)!j!}(\epsilon |\alpha|^2 )^{j}\langle\alpha|(\hat{M}_0^{\dagger})^{N-j}(\hat{M}_0)^{N-j}|\alpha\rangle\label{eq2}
\end{equation}
To leading order, we can approximate,
\begin{eqnarray}
    &&\langle\alpha|\{(\hat{M}_0^{\dagger})^{N-j}(\hat{M}_0)^{N-j}|\alpha\rangle\approx  \langle\alpha|\{(1-\epsilon a^{\dagger}a)^{N-j}\}|\alpha\rangle\nonumber\\&&\approx  \langle\alpha|(1-(N-j)\epsilon a^{\dagger}a)|\alpha\rangle\nonumber\approx 1-(N-j)\epsilon |\alpha|^2\approx (1-\epsilon |\alpha|^2)^{N-j}.
\end{eqnarray}
We therefore have,
\begin{equation}
    p(j,N) = \frac{N!}{(N-j)!j!}(\epsilon |\alpha|^2 )^{j}(1-\epsilon |\alpha|^2)^{N-j}.\label{EqB}
\end{equation}
We define $N\epsilon|\alpha|^2 = \lambda$. Then we can write,
\begin{equation}
    p(j,N) = \frac{N!}{(N-j)!j!}\bigg(\frac{\lambda}{N}\bigg)^{j}\bigg[1-\bigg(\frac{\lambda}{N}\bigg)\bigg]^{N-j} = \frac{\lambda^j}{j!}\frac{N!}{(N-j)!N^j}\bigg[1-\bigg(\frac{\lambda}{N}\bigg)\bigg]^{-j}\bigg[1-\bigg(\frac{\lambda}{N}\bigg)\bigg]^{N}.
\end{equation}
In the limit $N\rightarrow\infty$, we have,
\begin{equation}
    \frac{N!}{(N-j)!N^j}\rightarrow 1,~~\bigg[1-\bigg(\frac{\lambda}{N}\bigg)\bigg]^{-j}\rightarrow 1, ~~\text{and}~~\bigg[1-\bigg(\frac{\lambda}{N}\bigg)\bigg]^{N}\rightarrow e^{-\lambda}.
\end{equation}
Hence we obtain,
\begin{equation}
    \lim_{N\rightarrow\infty}p(j,N)\rightarrow \frac{\lambda^j}{j!}e^{-\lambda} = \frac{(N\gamma_0 \Delta t|\alpha|^2)^j}{j!}e^{-N\gamma_0 \Delta t |\alpha|^2} = \frac{(\gamma_0 T|\alpha|^2)^j}{j!}e^{-\gamma_0 T|\alpha|^2} =p(j,T). \label{palpha}
\end{equation}
The average number of ticks in the duration $T=N\Delta t$ is $\bar{j}=N\gamma_0 \Delta t |\alpha|^2 =\gamma_0 |\alpha|^2 T$. Now we can generalize Eqs.~\eqref{eq2}--\eqref{palpha}
using the diagonal $P$ representation for an arbitrary initial state of the field, as,
 \begin{equation}
    P(j,T)=\int d^2\beta P(\beta)\frac{1}{j!}[\gamma_0 |\beta|^2 T]^j e^{-\gamma_0 |\beta|^2 T},\label{eqlast}
\end{equation}
that agrees to the result in Ref.~\cite{Mandel_Wolf_1995}.


\section{The ratio test for a generic Gaussian state\label{arbGaus}}
In order to compute the ratio test for an arbitrary quantum mechanical state, we only need to evaluate the probability $P_{0}$, as the probabilities $P_{1}$ and $P_{2}$ can be derived from $P_{0}$ using Eq.~\eqref{ArbApprox} or Eq.~\eqref{pS}. Here we estimate $P_{0}$ for an arbitrary Gaussian state, which can be a displaced, squeezed, thermal state. The key insight that allows us this calculating this is that,
\begin{equation}
    P_{0} = \langle e^{-\gamma_0\Delta t a^{\dagger}a}\rangle_{\rho}=\frac{e^{\gamma_0\Delta t}}{e^{\gamma_0\Delta t}-1}\text{tr}[\rho\rho_{\text{th}}^{'}],
\end{equation}
where $\rho_{\text{th}}^{'}$ is the thermal state with effective temperature $\beta\hbar\omega = \gamma_{0}\Delta t$. The corresponding occupation number $n_{\text{th}}^{'} = (e^{\gamma_0\Delta t}-1)^{-1}$. Now in order to evaluate $P_{0}$ for a generic Gaussian state, we can use the Wigner representation of thermal states and the Weyl mapping, where,
\begin{equation}
   P_{0} = \frac{e^{\gamma_0\Delta t}}{e^{\gamma_0\Delta t}-1}\text{tr}[\rho\rho_{\text{th}}^{'}]=\frac{e^{\gamma_0\Delta t}}{e^{\gamma_0\Delta t}-1}2 \pi\int \mathcal{W}_{\rho}(x,y)\mathcal{W}_{\rho_{\text{th}}^{'}}(x,y)dx dy.
\end{equation}
For a generic Gaussian state, the Wigner function is given by,
\begin{equation}
    \mathcal{W}_{\rho}(x,y) =\frac{1}{2\pi\sqrt{|V_{\rho}|}} e^{-(\Vec{X}-\vec{X}_0)^{T}.V_{\rho}^{-1}.(\Vec{X}-\vec{X}_0)/2},
\end{equation}
where $\Vec{X}=\{x,p\}$ and $V_{\rho}$ is the covariance matrix, given by (assuming squeezing parameter $\zeta = re^{i\phi}$),
\begin{equation}
    V_{\rho} = \frac{2n_{\text{th}}+1}{2}\begin{bmatrix}-\cos (\phi) \sinh (2 r)+\cosh (2 r)&-\sin (\phi) \sinh (2 r)\\-\sin (\phi) \sinh (2 r)&\cosh (2 r)+\cos (\phi) \sinh (2 r)\end{bmatrix}.
\end{equation}
In our notation, the covariance matrix for $\rho_{\text{th}}^{'}$ has the form,
\begin{equation}
    V_{\rho_{\text{th}}^{'}} = \frac{2n_{\text{th}}^{'}+1}{2}\begin{bmatrix}1&0\\0&1\end{bmatrix}.
\end{equation}
Now what remains to be evaluated to compute $P_{0}$ is the overlap integral of two Gaussian Wigner functions. It can be evaluated to obtain (see for example, Ref.~\cite{duda_gaussian_2019}),
\begin{eqnarray}
   P_{0} &=& \frac{e^{\gamma_0\Delta t}}{e^{\gamma_0\Delta t}-1}\text{tr}[\rho\rho_{\text{th}}^{'}]=\frac{e^{\gamma_0\Delta t}}{e^{\gamma_0\Delta t}-1}2 \pi\int \mathcal{W}_{\rho}(x,y)\mathcal{W}_{\rho_{\text{th}}^{'}}(x,y)dx dy
   \nonumber\\
   &=&\frac{e^{\gamma_0\Delta t}\exp\bigg(-\frac{1}{2} \Vec{X}_0^T  V_{\rho}^{-1}\bigg( V_{\rho_{\text{th}}^{'}}^{-1}+ V_{\rho}^{-1}\bigg)^{-1} V_{\rho_{\text{th}}^{'}}^{-1}\Vec{X}_0\bigg)}{(e^{\gamma_0\Delta t}-1)\sqrt{|  V_{\rho_{\text{th}}^{'}}|  |  V_{\rho}|  |  V_{\rho_{\text{th}}^{'}}^{-1}+ V_{\rho}^{-1}| }}.
\end{eqnarray}
Now, for the null test, note $P_{1}$ and $P_{2}$ can be computed using the relations,
\begin{eqnarray}
P_{1} &=& -(\gamma_0\Delta t)\frac{d}{d(\gamma_0\Delta t)}P_{0},~~\text{and}~~P_{2} =\frac{(\gamma_0\Delta t)^2}{2}\bigg[\frac{d^2}{d(\gamma_0\Delta t)^2}+ \frac{d}{d(\gamma_0\Delta t)}\bigg]P_{0},
\end{eqnarray}
which follows from Eq.~\eqref{ArbApprox} or Eq.~\eqref{pS}. As an example, consider $\vec{X}=\{x_0,0\}$. In this case, we find, to leading order,
\begin{eqnarray}
    R&\approx& \frac{4 n_{\text{th}}^2-8 n_{\text{th}} x_{0}^2 \cos (\phi) \sinh (2 r)+8 (2 n_{\text{th}}+1) \left(x_{0}^2-1\right) \cosh (2 r)}{2 \left((2 n_{\text{th}}+1) \cosh (2 r)+x_{0}^2-1\right)^2}\nonumber\\&+&\frac{3 (2 n_{\text{th}}+1)^2 \cosh (4 r)+4 n_{\text{th}}-8 x_{0}^2 \cos (\phi) \sinh (r) \cosh (r)+2 x_{0}^4-8 x_{0}^2+5}{2 \left((2 n_{\text{th}}+1) \cosh (2 r)+x_{0}^2-1\right)^2}.
\end{eqnarray}
We see that, for a generic Gaussian state that is highly occupied, $1\leq R\leq 3+\text{cosech}^{2}(r)$ approximately, demonstrating the significant deviation of $R$ from one possible for generic Gaussian states. Coherent states and highly squeezed states are at opposite ends of the spectrum, with thermal states yielding $R=2$.
\end{widetext}
\bibliography{ref}
		   \end{document}